\documentclass[journal]{IEEEtran}
\usepackage{amsmath}
\usepackage{lineno,hyperref}
\modulolinenumbers[5]
\usepackage{amssymb}
\usepackage{booktabs}
\usepackage{graphicx}
\usepackage{bm}
\usepackage{cite}
\usepackage{multirow}

\usepackage[table,xcdraw]{xcolor}
\begin{document}
\title{MDA GAN: Adversarial-Learning-based 3-D Seismic Data Interpolation and Reconstruction for Complex Missing}
\author{Yimin Dou, Kewen Li, Hongjie Duan, Timing Li, Lin Dong, Zongchao Huang
	\thanks{The corresponding author is Kewen Li. likw@upc.edu.cn}
	\thanks{Yimin Dou, Kewen Li, Zongchao Huang, College of computer science and technology, China University of Petroleum (East China) Qingdao, China.}
	\thanks{Timing Li, College of Intelligence and Computing, Tianjin University Tianjin, China.}
	\thanks{Hongjie Duan, Shengli Oilfield Company, SINOPEC Dongying, China.}
	\thanks{Lin Dong, Center on Frontiers of Computing Studies, Peking University Beijing, China}
	\thanks{This work was supported by grants from the National Natural Science Foundation of China (Major Program, No.51991365), and the Natural Science Foundation of Shandong Province, China (ZR2021MF082).}
}

\maketitle
\begin{abstract}
The interpolation and reconstruction of missing traces is a crucial step in seismic data processing, moreover it is also a highly ill-posed problem, especially for complex cases such as high-ratio random discrete missing, continuous missing and missing in fault-rich or salt body surveys. These complex cases are rarely mentioned in current works.
To cope with complex missing cases, we propose Multi-Dimensional Adversarial GAN (MDA GAN), a novel 3-D GAN framework.
It keeps anisotropy and spatial continuity of the data after 3D complex missing reconstruction using three discriminators. The feature splicing module is designed and embedded in the generator to retain more information of the input data. The Tanh cross entropy (TCE) loss is derived, which provides the generator with the optimal reconstruction gradient to make the generated data smoother and continuous.
We experimentally verified the effectiveness of the individual components of the study and then tested the method on multiple publicly available data. 
The method achieves reasonable reconstructions for up to 95\% of random discrete missing and 100 traces of continuous missing.
In fault and salt body enriched surveys, MDA GAN still yields promising results for complex cases.
Experimentally it has been demonstrated that our method achieves better performance than other methods in both simple and complex cases. 
Moreover, our network does not require training weights for each survey, the same weights it uses are applied to multiple surveys, significantly reducing time and computational costs, and we make the model publicly available on https://github.com/douyimin/MDA\_GAN.

\end{abstract}


\IEEEpeerreviewmaketitle

\section{Introduction}
Complete seismic data are often difficult to acquire due to various constraints such as economic, physical, and other factors. Reconstructing missing seismic data is a critical and challenging task. There are currently two major types of methods for interpolation and reconstruction of seismic data, theory-driven and data-driven.

There are four theory-driven methods. The first are prediction filter-based methods\cite{spitz1991seismic,porsani1999seismic,gulunay2003seismic,naghizadeh2007multistep}, which extend interpolation to the frequency-space $(f-x)$ domain, which exploit the predictability of linear events in the $(f-x)$ domain. Such methods must use a local windowing strategy when handling nonlinear events. The choice of window affects the performance of interpolation and reconstruction, and the choice of the optimal window is currently unrealistic \cite{wang2019deep}. 
The second is the wave equation-based approach\cite{ronen1987wave, fomel2003seismic}, which requires subsurface velocity as a priori, however, it is difficult to obtain accurate velocity models in practical engineering, which limits its extension. 
The third is the methods based on sparse constraints \cite{sacchi1998interpolation,wang2010seismic,latif2016efficient,liu2015seismic}, which assume that the seismic data are linear or quasi-linear, and even though it can handle nonlinear data through parameter tuning, it is impractical to obtain optimal parameters, which affects its performance\cite{niu2021seismic}. 
Finally, there are low-order constraint methods\cite{oropeza2011simultaneous, huang2016damped, innocent2021robust,zhang2019nonconvex}, which are based on compression sensing, however most of them are only applicable to the missing case of random discrete.

Data driven employing machine learning or deep learning,  there are generally two types of methods, Auto-Encoder
(AE) and Generative Adversarial Neural Networks (GAN). AE-based methods include using AE\cite{wang2019deep}, CAE\cite{wang2020seismic}, UNet\cite{park2019reconstruction} and ResNet\cite{wang2018seismic}, etc. 
AE methods use encoders to extract hidden variables from missing data, and the complete data is used to supervise the seismic data generated by the decoder, which leads to data being learned point-to-point, which also results in less comprehensive global information acquisition, with weak global information and strong local information. Due to the lack of global information, the performance is weak for large scale missing, so most such methods are suitable for handling random discrete missing \cite{yu2021attention,li2021consecutively,he2021seismic,li2017prediction,li2018long,qian2021dtae}. Wu et al. introduced the attention mechanism to solve the weak global information problem of the AE method, and achieved promising results in the case of continuous missing \cite{li2021consecutively,yu2021attention}.
He et al. used a multi-stage UNet to step through the interpolation of consecutive missing \cite{he2021seismic}.

Additional to this are GAN-based methods, but they are less studied. GAN adds discriminator to AE (generator), which introduce regional or global information, so it should have better performance, especially for continuous and large scale deficiencies \cite{misra2019deep}.
Oliveira achieved interpolation and reconstruction of  Netherlands Offshore F3 seismic data via cGANs \cite{oliveira2018interpolating}. 
These methods all employ $L_1$ as the reconstruction loss of the generator, but GAN generally uses Tanh as the activation function of the output layer to ensure that the value domain of the generator is at $[-1,1]$, and the gradient of $L_1$ under this activation function is non-positively correlated with the loss (equations (\ref{l1}, \ref{l1_2}, \ref{3})), which may lead to point-to-point learning that cannot be optimized. 
Since natural images are bounded by the three RGB channels to each other, and allow for diversity inpainting, $L_1$ is possible as the reconstruction loss of natural image inpainting \cite{newell2016stacked,badrinarayanan2017segnet,ronneberger2015u,noh2015learning}. However, seismic data only has single channel, so these GAN-based interpolation results for seismic data have some obvious splicing traces, and the interpolation and reconstruction of seismic data require restoring the original seismic data as much as possible, and $L_1$ causes the results to deviate from the optimal solution.
Wei et al. changed the adversarial loss to Wasserstein loss based on cGAN and achieved 2-D interpolation of up to 35 consecutive seismic traces missing.\cite{wei2021aliased}. Kaur et al. used GAN and CycleGAN to interpolate 2-D synthetic seismic data \cite{kaur2019seismic,kaur2021seismic}.
The existing methods are limited to the application of GAN in the field of seismic interpolation, so there is no significant difference in the performance of the current GAN-based and the AE-based methods.
Moreover, most of the existing AE and GAN-based methods focus only on repairing 2D slices, but reconstruction 2D slices one by one leads to discontinuities in brutal stacked \cite{niu2021seismic} section when there are complex cases such as large scale or continuous missing in shot gathers. Therefore it is crucial to research 3D interpolation and reconstruction approaches.

We suggest that introducing GAN into seismic interpolation requires addressing two major challenges. The first is how to design a reasonable GAN benchmark framework for 3D seismic data interpolation and reconstruction to prevent mode collapse during training, while ensuring that the anisotropy and spatial continuity of 3D seismic data can be preserved even in the case of complex missingness. Second, it can ensure that the generator performs accurate point-to-point learning without distortion of each pixel even when the discriminator introduces global information and uses the Tanh activation function.

There is no GAN-based method for 3D seismic interpolation \cite{oliveira2018interpolating,wei2021aliased,kaur2019seismic,kaur2021seismic,qian2021dtae}. The interpolation of 3D seismic data needs to ensure the spatial continuity and reliability of the seismic data in three directions (timeline, inline, and crossline) while maintaining the anisotropy of the seismic data, especially when we need to handle complex problems such as large scale discrete missing or continuous missing. 
Moreover, compared with the reconstruction of 2D data, the computational resources and parameters required for 3D data have increased significantly\cite{wu2016learning,cirillo2020vox2vox}, especially the alternate training mode of generator and discriminator, which puts harsh requirements on the hardware conditions. The increase of parameters is more possible to cause the mode collapse of GAN \cite{radford2015unsupervised, gulrajani2017improved}, so it is challenging to design a GAN-based interpolation and reconstruction framework for 3D seismic data.

We propose a Multi-Dimensional Adversarial GAN (MDA GAN) that uses three discriminators to extract regional or global features for 3D volume, timeline, inline, and crossline of seismic data (inline and crossline share one discriminator and apply to both pre-stack and post-stack data) to maintain anisotropy and spatial continuity of the reconstructed seismic data.

In response to the demanding hardware requirements imposed by 3D data and the mode collapse problems that may be caused, a 3D data generator for GAN is designed, which propagates features using high and low resolution in parallel to prevent information loss, so that excessive width is not required to retain image information, and low (large scale) and high resolution interact during propagation, increasing the receptive field and fully extracting global information, so the network does not need to be deeper \cite{dou2021efficient}. Meanwhile, we propose the FSM module and embed this generator to ensure the integrity of the unmissed parts during the model inference. The generator reduces the parameters and the load on the hardware, while maintaining a high quality reconstruction performance.

To enable accurate point-to-point learning of seismic data in the framework of GAN, we analyzed the gradients of reconstruction loss that are commonly used nowadays, and thus derived the Tanh Cross Entropy (TCE) loss that can perform accurate pixel-level learning under Tanh activation function for seismic data reconstruction by MDA GAN.

In general, our contributions include the following three points.
Firstly, a novel GAN framework, MDA GAN, is proposed to cope with more complex 3D seismic missing cases to ensure the anisotropy as well as spatial continuity of the reconstructed 3D data.
Secondly, to cope with the burden on the hardware caused by the parameter proliferation brought by 3D GAN, we elaborated the generator and proposed and embedded the FSM to accomplish end-to-end data generation and splicing.
Finally, TCE reconstruction loss is proposed so that the training process can provide smoother gradients, enable accurate point-to-point learning, and ensure that the reconstructed data are distortion-free.

\section{Approach}
In this section, we first introduce the baseline framework of MDA GAN, then illustrate the structure of FSM and how it is embedded and functions in the generator, and finally discuss the seismic voxel (pixel) level reconstruction loss Tanh Cross Entropy Loss (TCE) and analyze its training gradient to illustrate the advancement of the method.

\subsection{Baseline Framework}
\begin{figure*}[htb]
	\centering
	\includegraphics[scale=0.70]{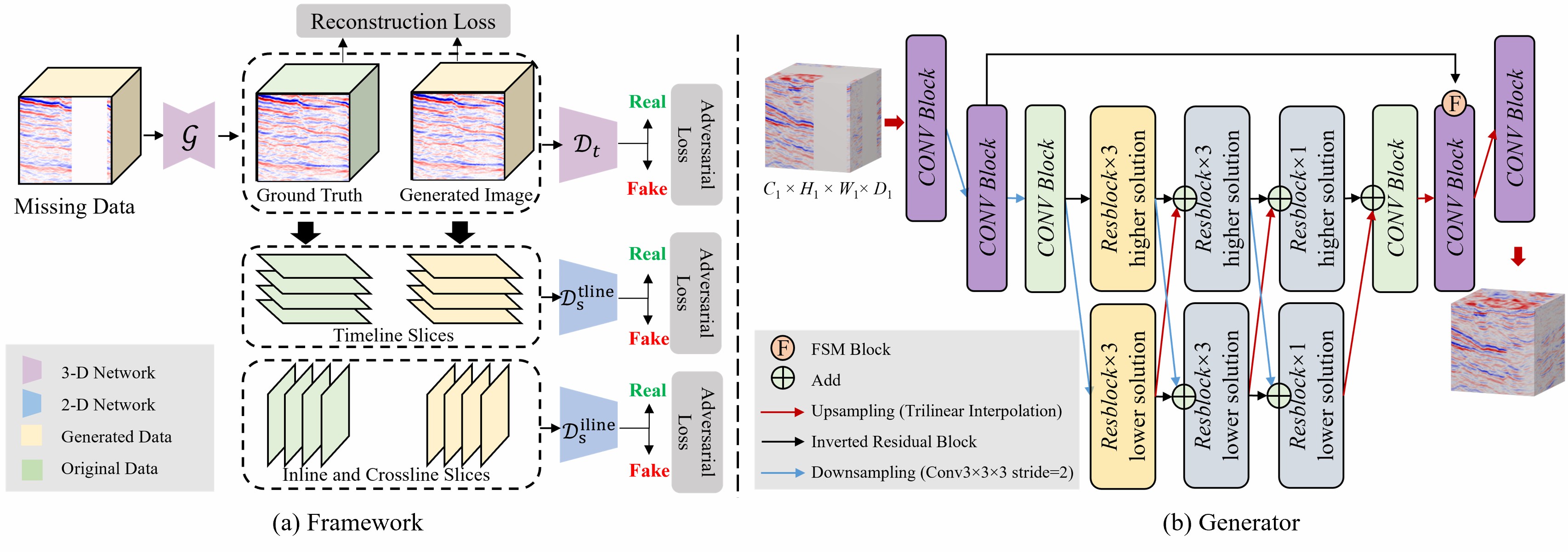}
	\centering\caption{In (a), the framework consists of one 3-D generator, one 3-D discriminator and two 2-D discriminators. For training, the input to the 3D network is data of size $128\times 128\times 128$, and the batch size is $b$. To conserve the RAM, the 2D discriminator randomly draws $8$ slices of  $128 \times 128$ in the 3D data along the corresponding direction as input, and the batch size is $8\times b$. While for inference, the input to the generator can be any size as allowed by the hardware. (b) is the detailed structure of the generator in the framework, and the discriminator follows the standard encoder structure.The CONV block consists of a $3\times3$ convolution, a normalization layer and a LeakyReLU activation function, Resblock was proposed by He et al \cite{he2016deep}.}
	\label{arch}
\end{figure*}
Our approach is based on GAN, and in \ref{appendix} Appendix we add a basic explanation for GAN and the current underlying framework for GAN-based image inpainting in computer vision.

The major disadvantage of current 2D seismic interpolation is that reconstruction 2D slices one by one leads to discontinuities in brutal stacked section\cite{niu2021seismic} when there are complex cases such as large scale or continuous missing in shot gathers. To solve this problem, we extend the seismic interpolation and reconstruction to three dimensions.

The generator is a 3D CNN structure, and the input and output are 3D data of the same size. This article describes the structure of the generator and its components detailed in \ref{gen_part}.

The 3D seismic data have obvious anisotropy, where the seismic reflection axis is approximately perpendicular to the inline and crossline slices, and a clear seismic layered texture can be observed in these two directions, while the timeline direction shows irregular texture. Therefore, we use two 2D discriminators  $\mathcal{D}^m_{\theta_{D}^\text{xi}}$ and $\mathcal{D}^m_{\theta_{D}^\text{t}}$, inline and crossline share discriminator $\mathcal{D}^m_{\theta_{D}^\text{xi}}$, and timeline direction uses $\mathcal{D}^m_{\theta_{D}^\text{t}}$ discriminator.
2D discriminators are employed for each directional slice to maintain the anisotropy of the restored seismic data, and we also use a 3D discriminator $\mathcal{D}^t_{\theta_{D}}$ to ensure the continuity of the results in 3D space.

The framework consists of a 3D generator, two 2D discriminators and a 3D discriminator. Accordingly, we use three adversarial losses $\mathcal{L}_{\mathcal{G}^t}$, $\mathcal{L}_{\mathcal{G}^m_\text{xi}}$ and $\mathcal{L}_{\mathcal{G}^m_\text{t}}$. In combination with the reconstruction (\ref{inpvg}), the loss of the supervised generator can be expressed as equation (\ref{adloss3}).

\begin{equation}
	\mathcal{L}_\mathcal{G} = \mathcal{L}_{\mathcal{G}^t} + \mathcal{L}_{\mathcal{G}^m_\text{xiline}} + \mathcal{L}_{\mathcal{G}^m_\text{tline}}+ \lambda\mathcal{L}_\text{rec} \label{adloss3}
\end{equation}
where $\mathcal{L}_\text{rec}$ is the reconstruction loss, which we will describe in detail in \ref{recloss}, $\lambda$ is a scaling factor to adjust the reconstruction loss to the same order of magnitude as the adversarial loss.
Each adversarial loss is expressed as equation (\ref{adloss2}).
\begin{equation}
	\mathcal{L}_\text{adv} = \text{log}(1 - \mathcal{D}_{\theta_{D}}(\mathcal{G}_{\theta_{G}}(\textbf{I}_m)))\label{adloss2}
\end{equation}
where $\textbf{I}_m$ is the missing input data. This loss tries to make the generator fool the discriminator so that the generated hidden variables are approximated to the real data.

Each discriminator determines the authenticity of the generated tensor or matrix, using the cross-entropy loss for supervision, equation (\ref{dloss5}).

\begin{equation}
	\mathcal{L}_\mathcal{D} = \frac{1}{2} \text{log}(1 - \mathcal{D}_{\theta_{D}}(\textbf{I}_g)) +\frac{1}{2}  \text{log}( - \mathcal{D}_{\theta_{D}}(\mathcal{G}_{\theta_{G}}(\textbf{I}_m))) \label{dloss5}
\end{equation}
For reconstructing high-quality seismic data, our approach employs adversarial learning in multiple dimensions, hence the term Multi-Dimensional Adversarial (MDA) GAN. The overall framework of MDA GAN is shown in Fig.\ref{arch}-(a).

\subsection{Generator} \label{gen_part}

\subsubsection{Backbone Network}
Traditional generators normally take the form of concatenating high-resolution features with low-resolution features, i.e., encoder-decoder structures, and related models are available for various vision tasks \cite{newell2016stacked,badrinarayanan2017segnet,ronneberger2015u,noh2015learning,dou2021attention}. However, such methods downsample the features several times and returns the original resolution via the decoder, which causes the loss of information, while the process of restoring the resolution requires plenty of parameters and eats up many video memory resources. The new model design idea of parallel propagation of high resolution with low resolution was proposed by Wang and Sun et al. \cite{sun2019deep,wang2020deep}, and our previous work demonstrated that this structured network can achieve higher performance with few hardware resources \cite{dou2021efficient}. We designed the generator based on this structure, using fewer computational resources to ensure the generation of high-quality data.

Resblock \cite{he2016deep} is selected as the base block of the generator. The network is downsampled only twice, and then a low-resolution branch is added for parallel propagation, which is structured as shown in Fig. \ref{arch}-(b). Because the network keeps the high-resolution features, we don't need an excessive width and depth to ensure that the network recovers from the low resolution, so we use fewer parameters and save much bandwidth.

\subsubsection{Feature Splicing Module}

To adequately preserve the unmissing information of the input data $\textbf{I}_m$, many works perform splicing of  $\textbf{I}_m$ with $\textbf{I}_c$ (output) via mask, and the splicing process is expressed as equation (\ref{msk}).
\begin{equation}
	\textbf{I}_g \approx \textbf{I}_g' = (1 - \mathcal{M}) \times \textbf{I}_c  +  \mathcal{M} \times \textbf{I}_m  \label{msk}
\end{equation}
Where $\textbf{I}_g$ is the ground truth and $\textbf{I}_g'$ is the spliced data, $\mathcal{M}$ is mask, which is a matrix of the same size as the inputs, with 0 indicating missing data and the opposite with 1. However, for seismic data, it is difficult to obtain masks by a single step (threshold method) and some manual intervention is needed to splice or introduce other algorithms.
Not splicing will cause the resulting $(1 - \mathcal{M}) \times \textbf{I}_c$ to differ from the original data $(1 - \mathcal{M}) \times \textbf{I}_g$ as shown in Fig. \ref{exp3} (a-5),(b-5),(c-5).
We wanna to find a solution to replace the role of $\mathcal{M}$ and to form end-to-end training and inference.



For this purpose, we presented FSM. Our approach is to splice the low-level features of the network (representing the original non-missing part) with the high-level features (representing the generated part) so that the network automatically completes this process, and then reconstructs the spliced features to output the result. The existing work such as UNet\cite{ronneberger2015u} which concatenates the high level and bottom level features and then fuses them using convolution does not express the splicing process explicitly, and in the fusion process, the missing and unmissing of the bottom level features but with have the same weight, this causes the generated unmissing part $(1 - \mathcal{M}) \times \textbf{I}_c$  to be blurred and differ from the original data, making it necessary to manually splice the generated with the original data.

The module can be represented by the following equations (\ref{fsm1}, \ref{fsm2}, \ref{fsm3}).
\begin{equation}
	\mathcal{F}^{\text{branch}} = \textbf{F}^\text{cat}(\mathcal{F}^l,\mathcal{F}^h)\label{fsm1}
\end{equation}
\begin{equation}
	\mathcal{W}_{m} = \sigma(\textbf{F}^\text{conv}(\textbf{F}^\text{conv}(\mathcal{F}^{\text{branch}}, 3, 1, 2C_1, C_1),1, 1,  C_1, 2C_1))\label{fsm2}
\end{equation}

\begin{equation}
	\mathcal{F}^{\text{spl}} = \textbf{F}^\text{conv}(\mathcal{W}_m \odot \mathcal{F}_\text{branch}, 1, 1, 2C_1, C_1)\label{fsm3}
\end{equation}
In equation (\ref{fsm1}) the features are concatenated with the high-level feature $\mathcal{F}^h$ and the low-level feature $\mathcal{F}^l$ to obtain $\mathcal{F}^\text{branch}$, and the channel after concatenation is changed from $C_1$ to $2\times C_1$. $\mathcal{F}^l$ uses the features after the first convolution and $\mathcal{F}^h$ uses the features before the last residual block, so they both keep the original feature resolution. $\mathcal{F}^l$ has not been downsampled and convolved multiple times, so it retains all the information of the original data, and $\mathcal{F}^h$ has been filled with the missing parts after the generator network.

We want to express the splicing process of the features explicitly, so that the network automatically expresses equation (\ref{msk}), we compress and weight $\mathcal{F}^\text{branch}$ by a convolution, then recover it to the original channel and restrict its value domain to $[0,1]$ by the sigmoid ($\sigma (x_i)=\frac{1}{1+\text{exp}(-x_i)}$) activation function to obtain $\mathcal{W}_m$, the above process is represented by equation (\ref{fsm2}).

\begin{figure}[htb]
	\centering
	\includegraphics[scale=0.2]{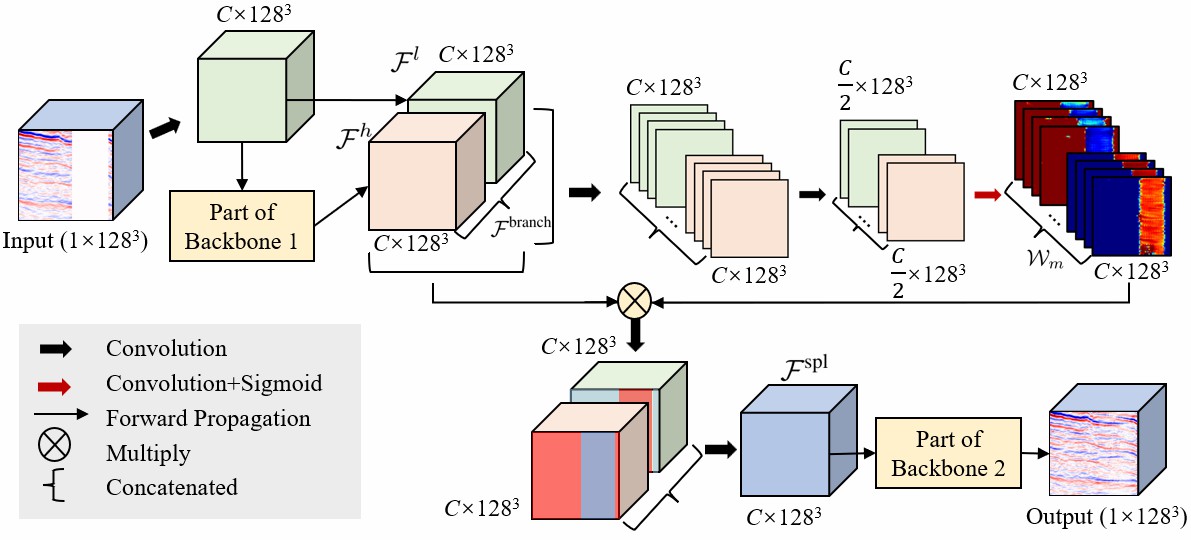}
	\centering\caption{The FSM workflow. The FSM selects the spatial response of the splicing by obtaining the high-dimensional mapping of $\mathcal{F}^\text{branch}$ to automate the splicing process.}
	\label{fsm}
\end{figure}

After equation (\ref{fsm3}), the FSM can be updated via standard backpropagation, and our experiments show that the module is able to converge $\mathcal{W}_m$ so that it generates a feature mask that resembles $\mathcal{M}$ and applies it to the splicing of features. The entire process can be represented in Fig. \ref{fsm}.

\begin{figure}[htb]
	\centering
	\includegraphics[scale=0.68]{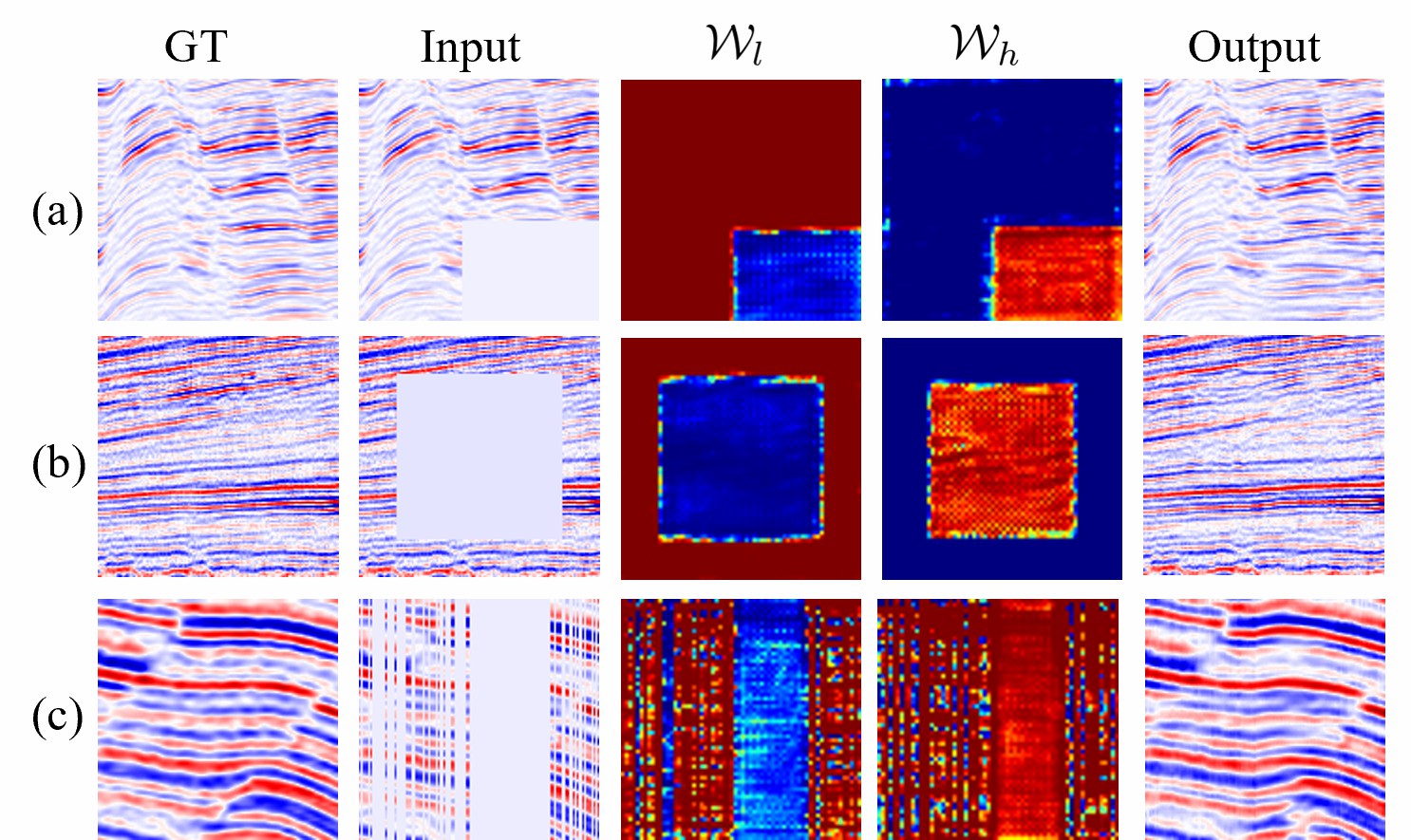}
	\centering\caption{The figure is shown as  2-D slices of $128^2$ in 3-D volumes of $128^3$, displaying the missing of the five modes. The FSM generates mask-like heatmaps without any mask supervision information.}
	\label{fmsre}
\end{figure}

We visualized $\mathcal{W}_m$, which has channel $2C_1$ and has two components, $\mathcal{W}_l$ ($C_1$ channels) and $\mathcal{W}_h$ ($C_1$ channels), with $\mathcal{W}_l$ corresponding to the $\mathcal{F}_l$ part of the $\mathcal{F}^\text{branch}$ and $\mathcal{W}_h$ corresponding to the $\mathcal{F}_h$ part, and Fig. \ref{fmsre} shows the weights of $\mathcal{W}_m$ with different patterns of input.

The approach don't input any supervised information about mask ($\mathcal{M}$), but the network still adaptively generates a mask-like heatmap. The supervision of the reconstruction loss and the FSM force $\mathcal{W}_m$ to automatically form semantic weights, and during continuous training, the network gradually learns which parts of the features should be extracted and retained.

\subsection{Discriminator}

The discriminator network uses the structure of a standard encoder. The discriminator consists of five convolutional layers, and each layer employs the LeakyReLU\cite{radford2015unsupervised} activation function, and also incorporates the spectral normalization\cite{miyato2018spectral} to stabilize the training of the GAN.

\subsection{Tanh Cross Entropy Loss} \label{recloss}

Most of the current reconstruction losses are based on  $L_1$ and $L_2$ \cite{pathak2016context,zheng2019pluralistic,yi2020contextual,zeng2021cr}, where $L_1$  provides stable gradients and $L_2$ has the more stable solution, but both loss functions have the common disadvantage of unsmoothed gradients in GANs (equations (\ref{3}) and (\ref{5})). Generator networks generally use Tanh as the final activation layer to keep the network's value domain at $[-1, 1]$, while ensuring training stability\cite{radford2015unsupervised}. Next we derive the gradient expressions for $L_1$ and $L_2$ under the Tanh activation function. 

\begin{figure}[htb]
	\centering
	\includegraphics[scale=0.29]{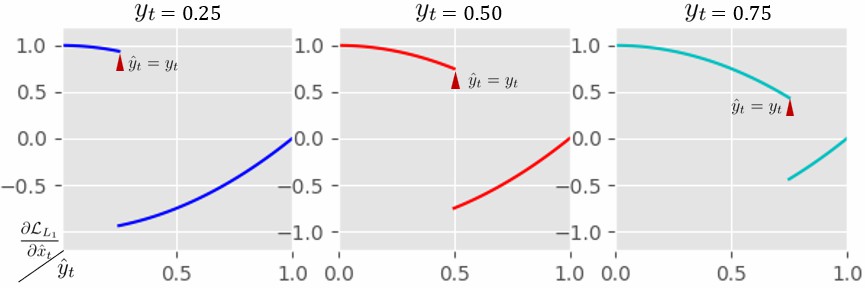}
	\centering\caption{Trend plot of $\partial\mathcal{L}_{L_1}/\partial\hat{x}_t$ with $\hat{y}_t$ for $L_1$ loss. This figure is from equation (\ref{3}).}
	\label{td1}
\end{figure}

First, we derive the gradient of $L_1$ loss back propagation to backbone.

\begin{equation}
	\mathcal{L}_{L_1} = \dfrac{1}{N_w\times N_h\times N_d} \sum_{i}^{N_w} \sum_{j}^{N_h} \sum_{k}^{N_d} \mid y_{i,j,k}-\hat{y}_{i,j,k}\mid_1 \label{l1}
\end{equation}

\begin{equation}
\begin{aligned}
\hat{y}_i = \text{Tanh}(\hat{x}) = \dfrac{e^{\hat{x}} - e^{-\hat{x}}}{e^{\hat{x}}+e^{-\hat{x}}}, \dfrac{d\hat{y}}{d\hat{x}} = 1 - \hat{y}^2
\end{aligned}\label{l1_2}
\end{equation}

\begin{equation}
\begin{aligned}
\dfrac{\partial\mathcal{L}_{L_1}(y,\hat{y})}{\partial\hat{x}_t} &= \dfrac{\partial\mathcal{L}_{L_1}(y,\hat{y})}{\partial\hat{y}_t}\cdot\dfrac{\partial\hat{y}_t}{\partial\hat{x}_t} \\
&=
\begin{cases}
	1-\hat{y}_t^2 \ \ \ \text{if} \ \ y_t-\hat{y}_t\ge0 \\
	\hat{y}_t^2-1 \ \ \ \text{if} \ \ y_t-\hat{y}_t<0 
\end{cases}
\end{aligned}\label{3}
\end{equation}
The most reasonable relationship between $\partial\mathcal{L}_{L_1}/\partial\hat{x}_t$ and $\hat{y}_t$ should be the one that triggers the minimum gradient (minimum absolute value) when $\hat{y}_t=y_t$. As $\hat{y}_t$ and $y_t$ become farther apart, the greater the gradient they cause should be, and this increasing relationship should be linear, i.e., smooth, its slope should not change with $\hat{y}_t$.
We define this optimal assumption as the equation (\ref{optima}).
\begin{equation}
\dfrac{\partial\mathcal{L}(y,\hat{y})}{\partial\hat{x}_t} = y_t - \hat{y}_t  = \varepsilon \label{optima} 
\end{equation}
Fig. \ref{td1} shows a sharp fluctuation in the neighborhood of $L_1$ loss when $\hat{y}_t=y_t$.
Although it provides the correct gradient direction, which leads to it being trainable, but its values are coarse and the incremental relationship is nonlinear, so it obviously does not satisfy the above optimality assumption.

Equations (\ref{l2}) and (\ref{5}) illustrate the gradient form of $L_2$ back propagation to backbone.

\begin{equation}
	\mathcal{L}_{L_2} =\dfrac{1}{N_w\times N_h\times N_d} \sum_{i}^{N_w} \sum_{j}^{N_h} \sum_{k}^{N_d} \Vert y_{i,j,k}-\hat{y}_{i,j,k} \Vert ^2_2 \label{l2}
\end{equation}
\begin{equation}
	\begin{aligned}
		\dfrac{\partial\mathcal{L}_{L_2}(y,\hat{y})}{\partial\hat{x}_t} &= \dfrac{\partial\mathcal{L}_{L_2}(y,\hat{y})}{\partial\hat{y}_t}\cdot\dfrac{\partial\hat{y}_t}{\partial\hat{x}_t} \\
		 &=2(y_t-\hat{y}_t)(1-\hat{y}_t^2) 
	\end{aligned}\label{5}
\end{equation}

\begin{figure}[htb]
	\centering
	\includegraphics[scale=0.29]{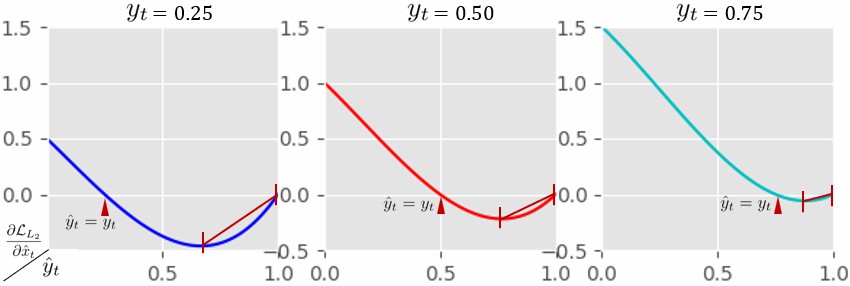}
	\centering\caption{Trend plot of $\partial\mathcal{L}_{L_2}/\partial\hat{x}_t$ with $\hat{y}_t$ for $L_2$ loss. This figure is from equation (\ref{5}).}
	\label{td2}
\end{figure}
In Fig. \ref{td2}, the three curves before reaching the extreme value point execute a relatively reasonable trend, although it is not smooth, the magnitude of their absolute values is proportional to the distance from the point $\hat{y}_t=y_t$. However, when they are in the position after the extreme point, i.e. the area marked by the red line segment, they show an unreasonable trend and the absolute values that should be large are becoming reduced. The $L_2$ loss forms a parabola between $\hat{y}_t=y_t$ and  $\hat{y}_t=1$. Although this can still be trained and does not affect the propagation of the gradient, we still hope to find a more reasonable gradient induced by the loss.

\begin{figure}[htb]
	\centering
	\includegraphics[scale=0.29]{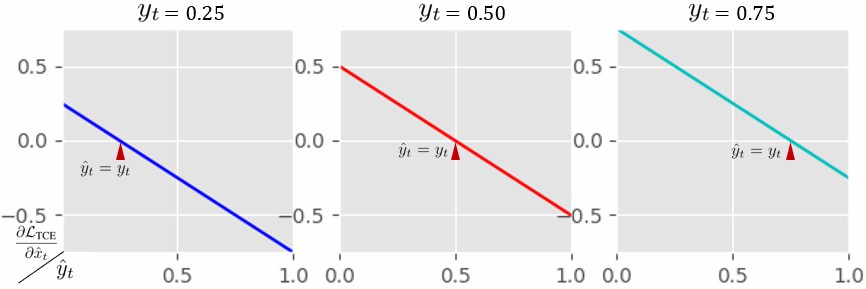}
	\centering\caption{Trend plot of $\partial\mathcal{L}_\text{TCE}/\partial\hat{x}_t$ with $\hat{y}_t$ for TCE loss. This figure is from equation (\ref{tce3}).}
	\label{td3}
\end{figure}

Both the $L_1$ and $L_2$ losses are deficient in the form of the gradient passed to the backbone. We give the optimal gradient form assumption in equation (\ref{optima}). 
We want  $\varepsilon$  to be directly used as the gradient form of the loss to backbone, and the resulting loss will satisfy the optimal gradient assumption, so we integrate it.
Then the loss function we expect can be expressed by the equation (\ref{tce1}).
\begin{small} 
\begin{equation}
	\begin{aligned}
	&\mathcal{L}_{\text{TCE}}(y_t,\hat{y}_t) =\int \varepsilon \cdot \dfrac{\partial\hat{x}_t}{\partial\hat{y}_t}d\hat{x}_t 
	=\int \dfrac{y_t-\hat{y}_t}{1-\hat{y}_t^2} d\hat{x}_t \\
	&=-y_t \left(\frac{\text{log}|\hat{y}_t+1|}{2} - \frac{\text{log}|\hat{y}_t-1|}{2} \right)\\
	&-\frac{1}{2}\text{log}|1-\hat{y}_t^2|+\text{log}2+C'\\
	&=-\left[ \left(\frac{1+y_t}{2} \right) \text{log}\left(\frac{1+\hat{y}_t}{2} \right) 
	+\left(\frac{1-y_t}{2} \right) \text{log}\left(\frac{1-\hat{y}_t}{2} \right)
	 \right]
	\end{aligned}\label{tce1}
\end{equation}
\end{small}
The loss of all voxels should be averaged during training, so that the loss expression is (\ref{tce2}).
\begin{small} 
\begin{equation}
	\begin{aligned}
	&\mathcal{L}_{\text{TCE}}(y,\hat{y}) = \dfrac{1}{N_w\times N_h\times N_d} \sum_{i}^{N_w} \sum_{j}^{N_h} \sum_{k}^{N_d} \\
	&\left[ \left(\frac{1+y_{i,j,k}}{2} \right) \text{log}\left(\frac{1+\hat{y}_{i,j,k}}{2} \right) +\left(\frac{1-y_{i,j,k}}{2} \right) \text{log}\left(\frac{1-\hat{y}_{i,j,k}}{2} \right)
	\right]
\end{aligned}\label{tce2}
\end{equation}
\end{small}
Its form is similar to cross-entropy, so we named it Tanh Cross-Entropy (TCE). This loss causes the gradient at training as in equation (\ref{tce3}).
\begin{equation}
	\dfrac{\partial\mathcal{L}_{\text{TCE}}(y,\hat{y})}{\partial\hat{x}_t} = y_t - \hat{y}_t  = \varepsilon \label{tce3}
\end{equation}
In Fig. \ref{td3}, TCE shows the smoothed gradient of its feedback to backbone. When $\hat{y}_t=y_t$, the gradient is $0$ and the distance between $\hat{y}_t$ and $y_t$ is linear with the gradient size, which is the optimal loss we want to obtain.

Although $L_1$ and $L_2$ can give correct feedback to the generator in the gradient direction, their gradient values are coarse and they do not reasonably reflect the distance between GT and the predicted value, while the gradient passed from TCE to the generator backbone is the residual of GT and the predicted value, which perfectly captures the change of the distance between the output and the target.

\subsection{Training}


The over-arching training framework follows the Fig. \ref{arch}. The optimizer uses Adam\cite{kingma2014adam}.
In general, the training of the discriminator should be ahead of the generator, which is more conducive to producing high-quality data. We use the TTUR strategy for training \cite{heusel2017gans}, setting the learning rate of discriminators $\eta_d = 0.0004$ and the generator $\eta_g = 0.0001$, finally alternating the training of the generator and discriminators.

\section{Experiments}
\subsection{Experimental Grouping}
We designed three control groups and one experimental group using the current most common UNet as the baseline model (Group 1), UNet has excellent performance in interpolating seismic data, which has been demonstrated in several works (\cite{park2019reconstruction,yu2021attention,he2021seismic} etc.).
Group 2 is the replacement of the backbone of Group 1 with the generator we proposed in Fig. \ref{arch}. Group 2 does not apply adversarial learning and therefore does not require tanh activation function and TCE loss, this group can verify the validity of the generator and the FSM compared to Group 1.
Group 3 incorporates multidimensional adversarial learning, which differs from the experimental group in that it uses $L_1$ as the reconstruction loss. This group can be compared with Group 2 to verify the effectiveness of multidimensional adversarial and with the experimental group to verify the effectiveness of TCE loss.
Group 4 is the experimental group, which uses exactly the method proposed in this paper.
The settings of the four groups of experiments are shown in Table  \ref{group}.
\begin{table}[htb]
	\centering
	\caption{Experimental grouping}
	\label{group}
	\begin{tabular}{@{}ccc@{}}
		\toprule
		& \textit{\textbf{Method}}       & \textit{\textbf{Explain}}                       \\ \midrule		                   
		Group 1 & UNet         & UNet-based regression, baseline model        \\
		Group 2 & MDA Generator      & Regression by generators only \\
		Group 3 & MDA GAN ($L_1$) & without TCE, with $L_1$ loss\\ 
		Group 4 & MDA GAN (TCE)    & Our method \\         \bottomrule
	\end{tabular}
\end{table}

\subsection{Dataset}
The current experimental approach of most data-driven seismic interpolation studies is to train a new set of weights for each survey \cite{wang2019deep,wang2020seismic,park2019reconstruction,wang2018seismic,yu2021attention,li2021consecutively,he2021seismic,qian2021dtae,he2021seismic,oliveira2018interpolating,kaur2019seismic,kaur2021seismic}, but retraining the neural network is extremely time-consuming and resource-wasting, resulting in a much higher time and resource cost than the theory-driven approach. 
Furthermore, the traces of the gathers are often highly similar to each other, so the neural network weights obtained from training for a specific survey may be overfitted, and this overfitting tends to show extremely high performance for the test set divided by the same survey, so this way of dividing the data set may not objectively reflect the performance of the model.

In this study, we want to train a general model that can be applied to the reconstruction of most seismic data without having to retrain new weights for different surveys, which is the current trend in AI development, CV and NLP scientists are currently working on building general models for various tasks\cite{dosovitskiy2020image, devlin2018bert}. We have used multiple surveys (both pre-stack and post-stack) to jointly train one model in the expectation that it will generalize to more data, we have made the final model publicly available on Github\footnote{Available: https://github.com/douyimin/MDA\_GAN}.

The usage of the data is shown in Table \ref{datausage}.
\begin{table}[htb]
	\caption{Data Usage}
		\centering
	\label{datausage}
	\begin{tabular}{@{}cc@{}}
		\toprule
		\textit{\textbf{Surveys}}                & \textit{\textbf{Purpose}}            \\ \midrule
		SEG C3                          & 3/5 train, 1/5 val, 1/5 test \\
		Mobil Avo Viking Graben Line 12 & 3/5 train, 1/5 val, 1/5 test \\
		F3 Netherlands                  & 3/5 train, 1/5 val, 1/5 test \\
		New Zealand Kerry               & 3/5 train, 1/5 val, 1/5 test \\
		New Zealand Parihaka            & 3/5 train, 1/5 val, 1/5 test \\
		New Zealand Opunake             & All train                   \\
		Sinopec surveys                 & All train                   \\ \bottomrule
	\end{tabular}
\end{table}

All training data are cropped to $128\times 128\times 128$, because the generator is fully convolutional, so the validation and test sets can be of arbitrary size. During training, 30\%-99\% of slices are randomly removed in the training cuboid as input data, the original data are used as supervised information, and the whole process does not require manual annotation. The training batch size is 4, and it is trained on two RTX3090Ti with 500000 steps. The model with the highest SSIM (details in \ref{metric}) metric on the validation set was selected as the final model and tested.

Next, we present and analyze the qualitative and quantitative results for the test set in each survey separately, and then present the combined quantitative metrics on the test set for all surveys. The trace of test data deletion in this paper's experiments were all performed using python's random library, and all random seeds were '2022'.
\subsection{Evaluation Metrics}\label{metric}
All experiments in this paper were evaluated quantitatively using Peak Signal-to-Noise Ratio (PSNR) and Structural Similarity (SSIM) metrics\cite{wang2004image}, and all data were normalized to $[0,1]$ prior to evaluation. To facilitate comparison and reproduction, the quantitative metrics are calculated directly using the public library scikit-image's implementation\footnote{Available: https://scikit-image.org} of PSNR and SSIM, and all hyperparameters in SSIM are default.
\subsubsection{PNSR}
PSNR is one of the most common methods for image quality assessment and is expressed as equation (\ref{psnr}).
\begin{equation}
	 \textbf{\text{F}}_\text{PNSR}(\textbf{I}_g', \textbf{I}_g) = 10\cdot \text{log}_{10}\left(\frac{\textbf{\text{F}}_\text{max}(\textbf{I}_g)^2}{\textbf{\text{F}}_\text{mse}(\textbf{I}_g', \textbf{I}_g) } \right)
	 \label{psnr}
\end{equation}
where $\textbf{\text{F}}_\text{max}(\cdot)$ is a function of the maximum value from the data, and $\textbf{\text{F}}_\text{mse}(\cdot)$ is the Mean Square Error (MSE) of the two images.
\subsubsection{SSIM}
Unlike PSNR's estimation of pixel point-to-point, SSIM incorporates consideration of inter-pixel dependence and prefers to evaluate the spatial similarity of two images. The expression is as follows.
\begin{equation}
	\textbf{\text{F}}_\text{SSIM}(\textbf{I}_g', \textbf{I}_g) = \frac{(2\mu_{\textbf{I}_g'}\mu_{\textbf{I}_g}+c_1)(2\sigma_{\textbf{I}_g'\textbf{I}_g}+c_2)}{(\mu_{\textbf{I}_g'}^2+\mu_{\textbf{I}_g}^2+c_1)(\sigma_{\textbf{I}_g'}^2+\sigma_{\textbf{I}_g}^2+c_2)}
	\label{ssim}
\end{equation}
Where $\mu_{\textbf{I}_g'}$ is the mean of $\textbf{I}_g'$, $\mu_{\textbf{I}_g}$ is the mean of $\textbf{I}_g$,  $\mu_{\textbf{I}_g'}^2$ is the variance of  $\textbf{I}_g'$, $\mu_{\textbf{I}_g}^2$ is the variance of  $\textbf{I}_g$, $\sigma_{\textbf{I}_g'\textbf{I}_g}$ is the covariance of  $\textbf{I}_g'$ and $\textbf{I}_g$, $c_1$ and $c_2$ two variables to stabilize the division with weak denominator, see literature \cite{wang2004image} for details.

In skit-image, the equation is implemented by a sliding window of size $7\times7\times7$, i.e., the local SSIM is calculated within each window to obtain the SSIM map. By averaging the SSIM map is the final SSIM value, while the visualization of the SSIM map can reflect the spatial distribution of the difference between the two data. In our experiments, we used SSIM to quantify the experimental results, while visualizing the SSIM map to clearly reflect the spatial distribution of the reconstructed differences.
\subsection{Test on SEG C3}
In SEG C3, we stacks gather into 3D, as shown in Table \ref{datausage} 3/5 of the data were involved in training, 1/5 were used for validation, and we used the last 1/5 for testing. Certainly, there are other surveys involved in the training.

\begin{figure}[htb]
	\centering
	\includegraphics[scale=0.45]{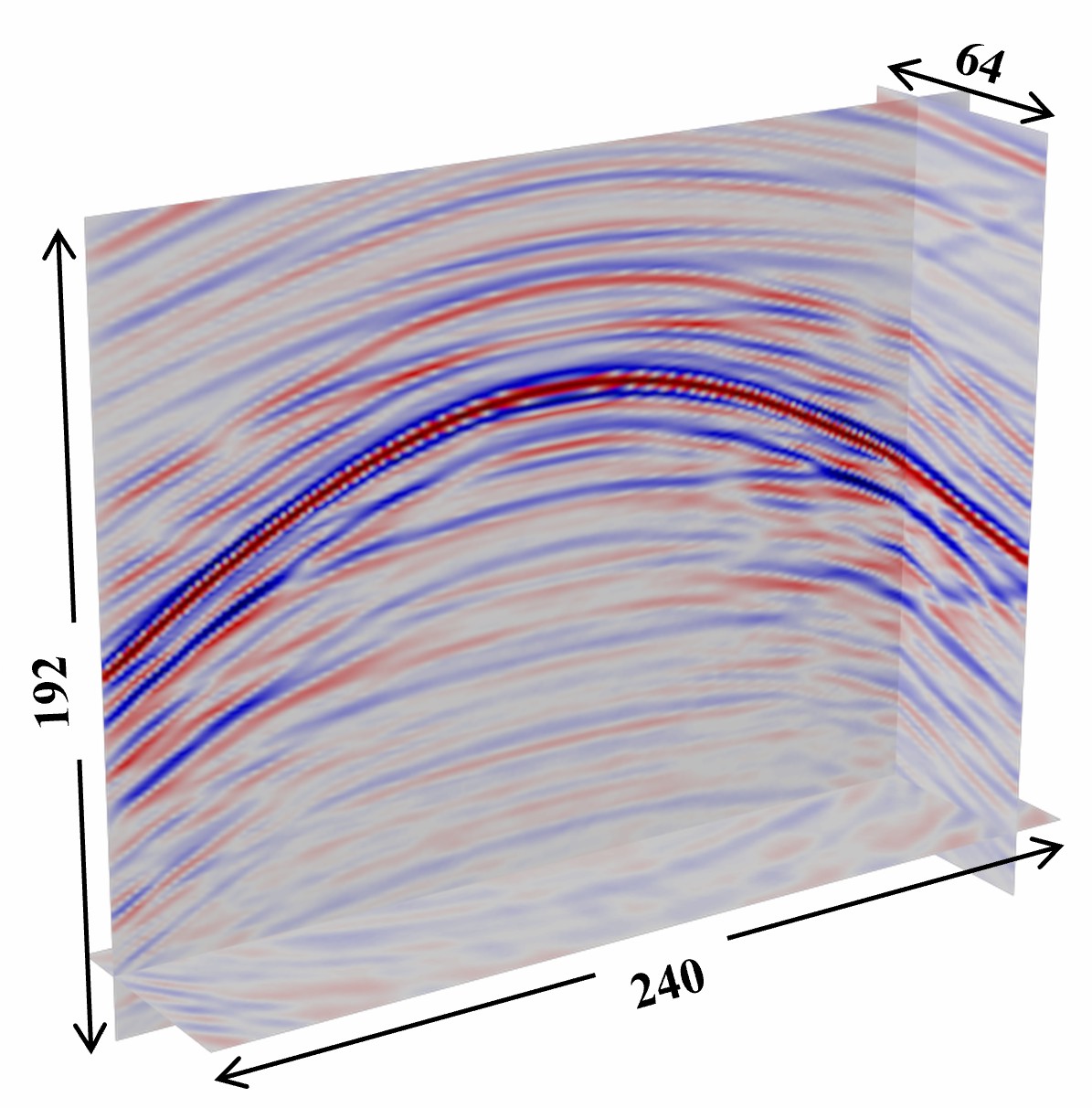}
	\centering\caption{SEG C3 test data.}
	\label{hti}
\end{figure}

We present it using one of the  $192\times 240\times 64$  grid points (Fig. \ref{hti}), which is obtained by interpolating (resize) the original data $168\times 205\times 31$.

\subsubsection{Discrete missing}
\begin{figure*}
	\centering
	\includegraphics[scale=0.52]{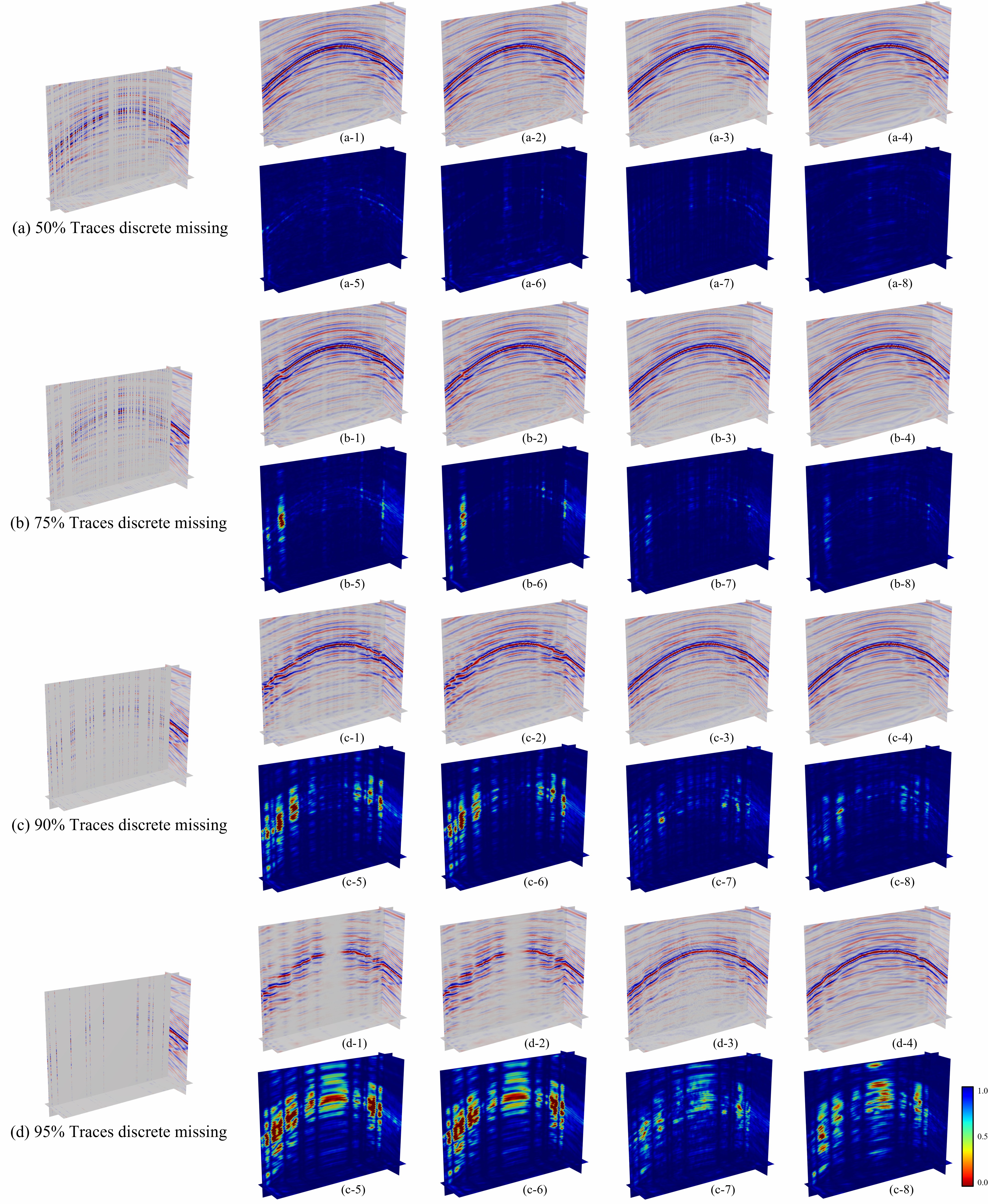}
	\centering\caption{(a) is the 50\% discrete missing case. (a-1)-(a-4) are the interpolation results of UNet, MDA Generator, MDA GAN ($L_1$) and MDA GAN methods for this case, respectively, and (a-5)-(a-8) are the SSIM Map corresponding to each method.
	(b) is the 75\% discrete missing case. (b-1)-(b-4) are the interpolation results of UNet, MDA Generator, MDA GAN ($L_1$) and MDA GAN methods for this case, respectively, and (b-5)-(b-8) are the SSIM Map corresponding to each method.
	(c) is the 90\% discrete missing case. (c-1)-(c-4) are the interpolation results of UNet, MDA Generator, MDA GAN ($L_1$) and MDA GAN methods for this case, respectively, and (c-5)-(c-8) are the SSIM Map corresponding to each method.
	(d) is the 95\% discrete missing case. (d-1)-(d-4) are the interpolation results of UNet, MDA Generator, MDA GAN ($L_1$) and MDA GAN methods for this case, respectively, and (d-5)-(d-8) are the SSIM Map corresponding to each method.}
	\label{exp1}
\end{figure*}

The upper limit of missing in most of the current studies is 50\%. The present work hopes to address more complex missing cases, and we use 50\% as the baseline and gradually expand the percentage of missing to 95\% (50\%, 75\%, 90\%, 95\%).
Fig. \ref{exp1} shows the qualitative results for each group of experiments in the missing randomized discrete case, this figure also illustrates the visualization of quantitative results using SSIM metrics. Table \ref{tab:exp1} shows the SSIM and PSNR values on this test data.

In the case of 50\% random missing, all four groups of experiments demonstrated promising results, and when the percentage of missing was expanded to 75\%, UNet's interpolation results started to show a discontinuity and its SSIM map showed a clear region of discrepancy, which was also shown at the same position in Group 2. When the percentage of missing continues to expand to 90\%, the reliable interpolation is completed only for groups 3 and 4 using multidimensional adversarial, and the regression-based approach is difficult to work with a high level of missingness.

With an extremely high percentage of missing 95\%, regression methods like Group 1 and 2 did not work at all, and interpolation was accomplished using multidimensional adversarial Group 3 and 4, which differed by using different reconstruction losses. Qualitatively, the results are smoother using TCE. Quantitatively, the method using $L_1$ has higher SSIM values and TCE shows a higher PSNR. From Fig. \ref{exp1} (a-3) (b-3) (c-3) and (d-3), $L_1$ also shows some discontinuities and distortions in other missing ratios, the phenomenon that is also present in many  GAN-based works\cite{kaur2021seismic,oliveira2018interpolating}. As we analyzed in \ref{recloss} section, the gradient provided by $L_1$ to the GAN is not optimal, which leads to interpolation results with some distortion. This is compensated well by TCE, which outperforms the currently most commonly used $L_1$ loss in most qualitative and quantitative experiments (except the SSIM metric under 95\% missing).

\begin{table}[htb]
	\centering
	\caption{SSIM and PSNR on SEG C3, Discrete Missing}
	\label{tab:exp1}
	\begin{tabular}{@{}cccccc@{}}
	\toprule
	\textit{\textbf{}} & \textit{\textbf{\begin{tabular}[c]{@{}c@{}}Discrete\\  Missing Ratio\end{tabular}}} & \textit{\textbf{50\%}} & \textit{\textbf{75\%}} & \textit{\textbf{90\%}} & \textit{\textbf{95\%}} \\ \midrule
	\textit{\textbf{}}     & UNet            & 0.9739 & 0.9356 & 0.8599 & 0.8189 \\
	\textit{\textbf{SSIM}} & MDA Generator   & 0.9711 & 0.9490 & 0.8803 & 0.8377 \\
	\textit{\textbf{}}     & MDA GAN ($L_1$) & 0.9772 & 0.9612 & 0.9474 & 0.8901 \\
	\textit{\textbf{}}     & MDA GAN (TCE)   & 0.9847 & 0.9651 & 0.9465 & 0.8854 \\ \midrule
	\textit{\textbf{}}     & UNet            & 36.88  & 31.20  & 26.05  & 22.22  \\
	\textit{\textbf{PSNR}} & MDA Generator   & 36.69  & 31.79  & 26.19  & 22.66  \\
	\textit{\textbf{}}     & MDA GAN ($L_1$) & 37.02  & 32.55  & 30.21  & 25.96  \\
	\textit{\textbf{}}     & MDA GAN (TCE)   & 37.14  & 32.81  & 31.00  & 26.29  \\ \bottomrule
\end{tabular}
\end{table}

\begin{figure*}
	\centering
	\includegraphics[scale=0.52]{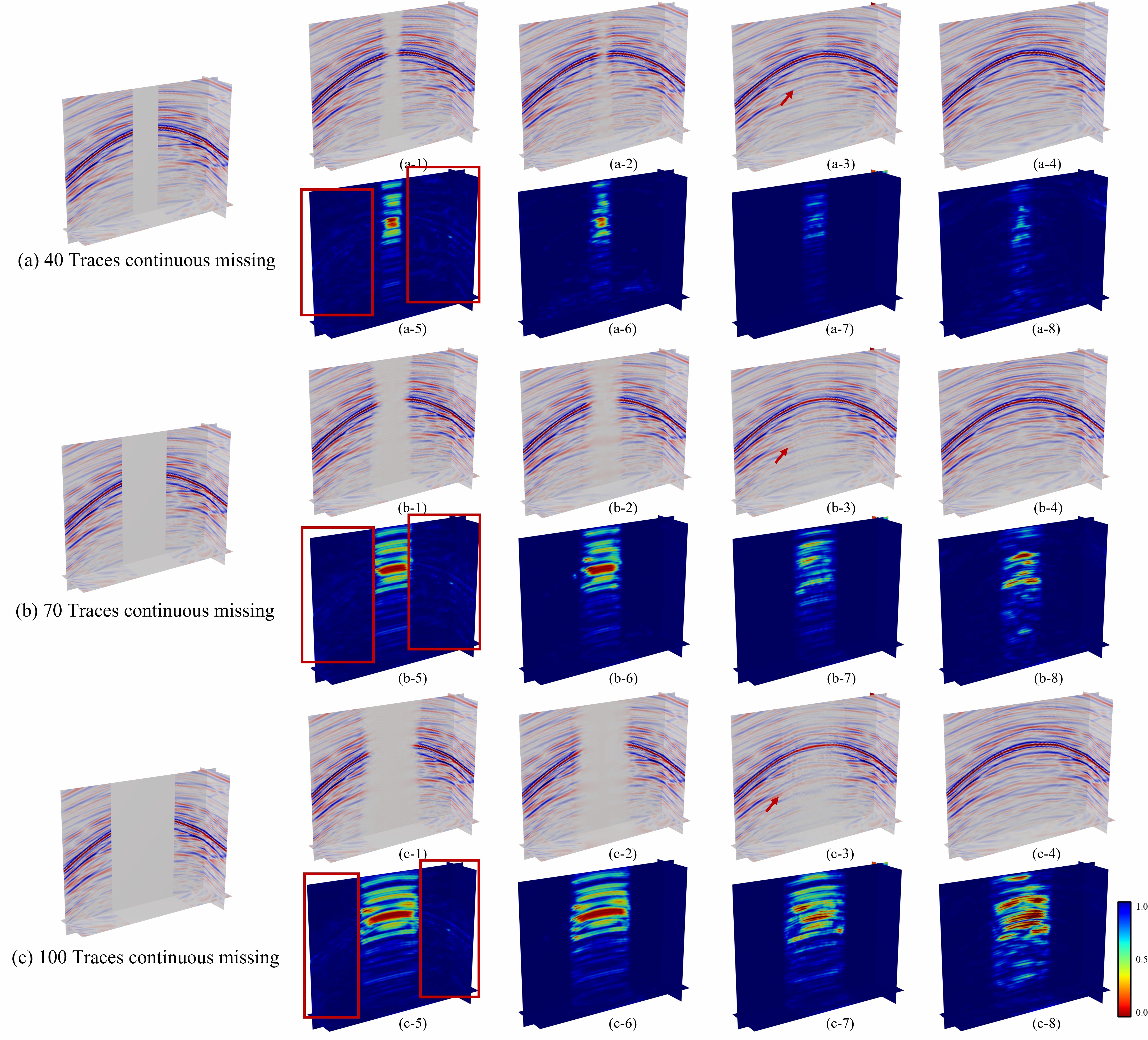}
	\centering\caption{	
		(a) is the case of 40 consecutive missing traces. (a-1)-(a-4) are the interpolation results of UNet, MDA Generator, MDA GAN ($L_1$) and MDA GAN methods for this case, respectively, and (a-5)-(a-8) are the SSIM Map corresponding to each method.
		(b) is the case of 70 consecutive missing traces. (b-1)-(b-4) are the interpolation results of UNet, MDA Generator, MDA GAN ($L_1$) and MDA GAN methods for this case, respectively, and (b-5)-(b-8) are the SSIM Map corresponding to each method.
		(c) is the case of 100 consecutive missing traces. (c-1)-(c-4) are the interpolation results of UNet, MDA Generator, MDA GAN ($L_1$) and MDA GAN methods for this case, respectively, and (c-5)-(c-8) are the SSIM Map corresponding to each method.}
	\label{exp3}
\end{figure*}

\subsubsection{Continuous missing}
Next, we compare the performance of the four methods in the continuous missing case. It should be noted that the best repair record for missing continuous seismic traces is currently by Yu et al.\cite{yu2021attention}. They interpolated $30\%$ of the missing continuous traces of the $128\times128$ 2-D seismic image with an interpolated area of $128\times38$, i.e. 38 traces.

Fig. \ref{exp3}  illustrates three consecutive missing, 40, 70, and 100 traces.
In the case of consecutive missing cases, Group 1 and 2 show unsatisfactory reconstructions with significant voids, and it can be observed that the regression method is only an extension of the data in the neighbourhood of the missing site and is limited in scope. Table \ref{tab:exp2} also shows that the quantitative metrics based on the MDA method are significantly higher than the other methods.

The MDA GAN method completes the reconstruction of the continuous missing. This experiment also shows the effectiveness of the TCE, where the interpolation result of $L_1$ has obvious splicing traces and discontinuity at the red arrow, while the method using TCE is smoother and continuous.

It is worth noting that the reconstruction results of UNet at the red box in Fig. \ref{exp3}  show significant differences compared to ground truth, while the model using FSM shows less or even no differences, precisely because the FSM in the generator comes into action, and as described in section \ref{gen_part} and Fig. \ref{fmsre}, FSM is able to retain more information about the unmissing part $(1 - \mathcal{M}) \times \textbf{I}_c$.

\begin{table}[htb]
	\centering
	\caption{SSIM and PSNR on SEG C3, Continuous Missing}
	\label{tab:exp2}
	\begin{tabular}{@{}ccccc@{}}
	\toprule
	\textit{\textbf{}} & \textit{\textbf{\begin{tabular}[c]{@{}c@{}}Continuous \\ Missing Traces\end{tabular}}} & \textit{\textbf{40}} & \textit{\textbf{70}} & \textit{\textbf{100}} \\ \midrule
	\textit{\textbf{}}     & UNet            & 0.9419 & 0.9122 & 0.8833 \\
	\textit{\textbf{SSIM}} & MDA Generator   & 0.9522 & 0.9248 & 0.9007 \\
	\textit{\textbf{}}     & MDA GAN ($L_1$) & 0.9799 & 0.9552 & 0.9225 \\
	\textit{\textbf{}}     & MDA GAN (TCE)   & 0.9789 & 0.9596 & 0.9188 \\ \midrule
	\textit{\textbf{}}     & UNet            & 30.18  & 25.94  & 22.35  \\
	\textit{\textbf{PSNR}} & MDA Generator   & 32.00  & 27.77  & 23.63  \\
	\textit{\textbf{}}     & MDA GAN ($L_1$) & 34.60  & 30.81  & 26.90  \\
	\textit{\textbf{}}     & MDA GAN (TCE)   & 34.78  & 31.11  & 26.85  \\ \bottomrule
\end{tabular}
\end{table}

\subsection{Test on Mobil Avo Viking Graben Line 12}
\begin{figure}[htb]
	\centering
	\includegraphics[scale=0.5]{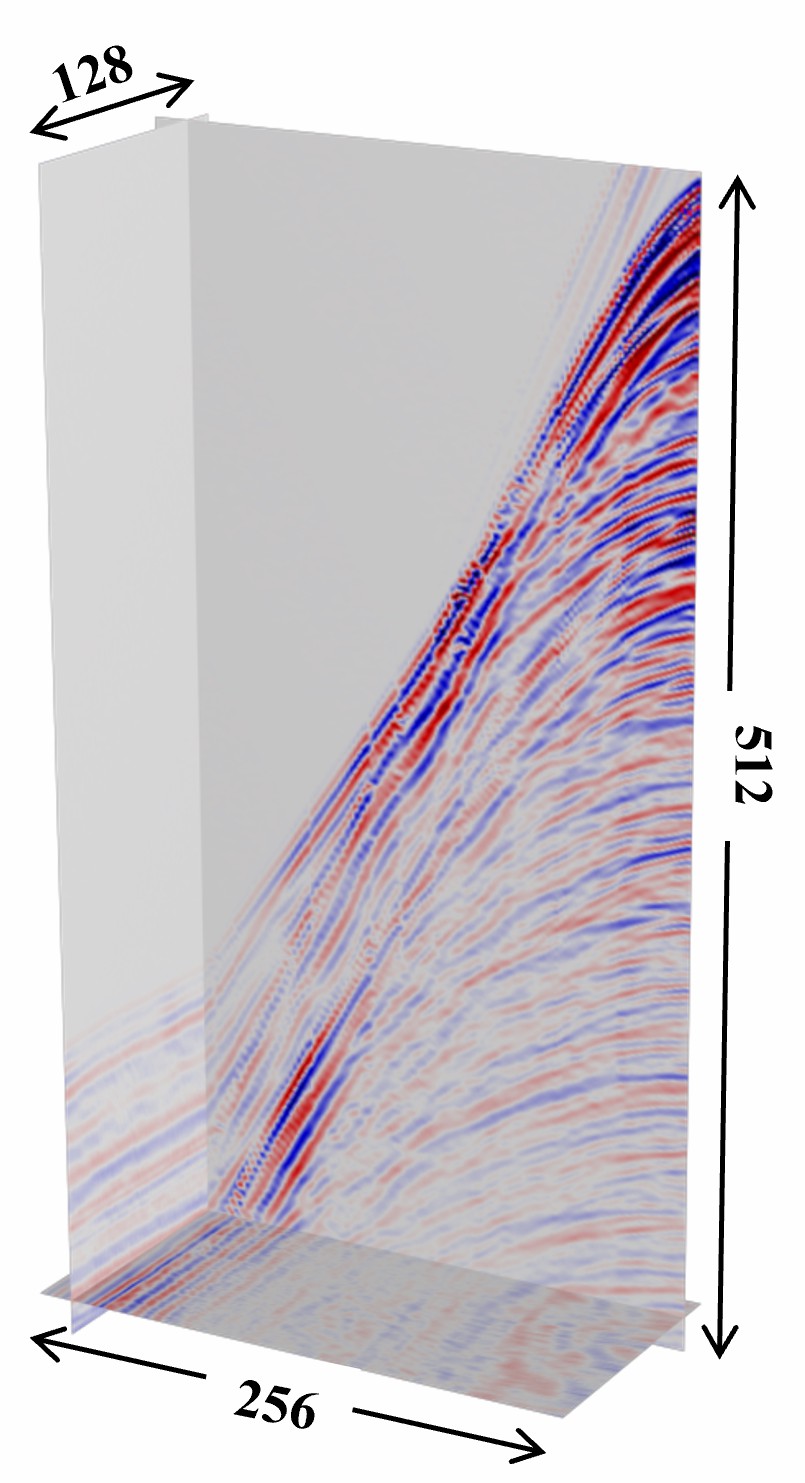}
	\centering\caption{
	Mobil Avo Viking Graben Line 12 test data
	}
	\label{mobil}
\end{figure}

We fold the prestack gather into 3D, and intercept $512\times 256\times 128$ grid points in the test data as a sample for qualitative analysis, as shown in Fig. \ref{mobil}.

\begin{figure*}
	\centering
	\includegraphics[scale=0.62]{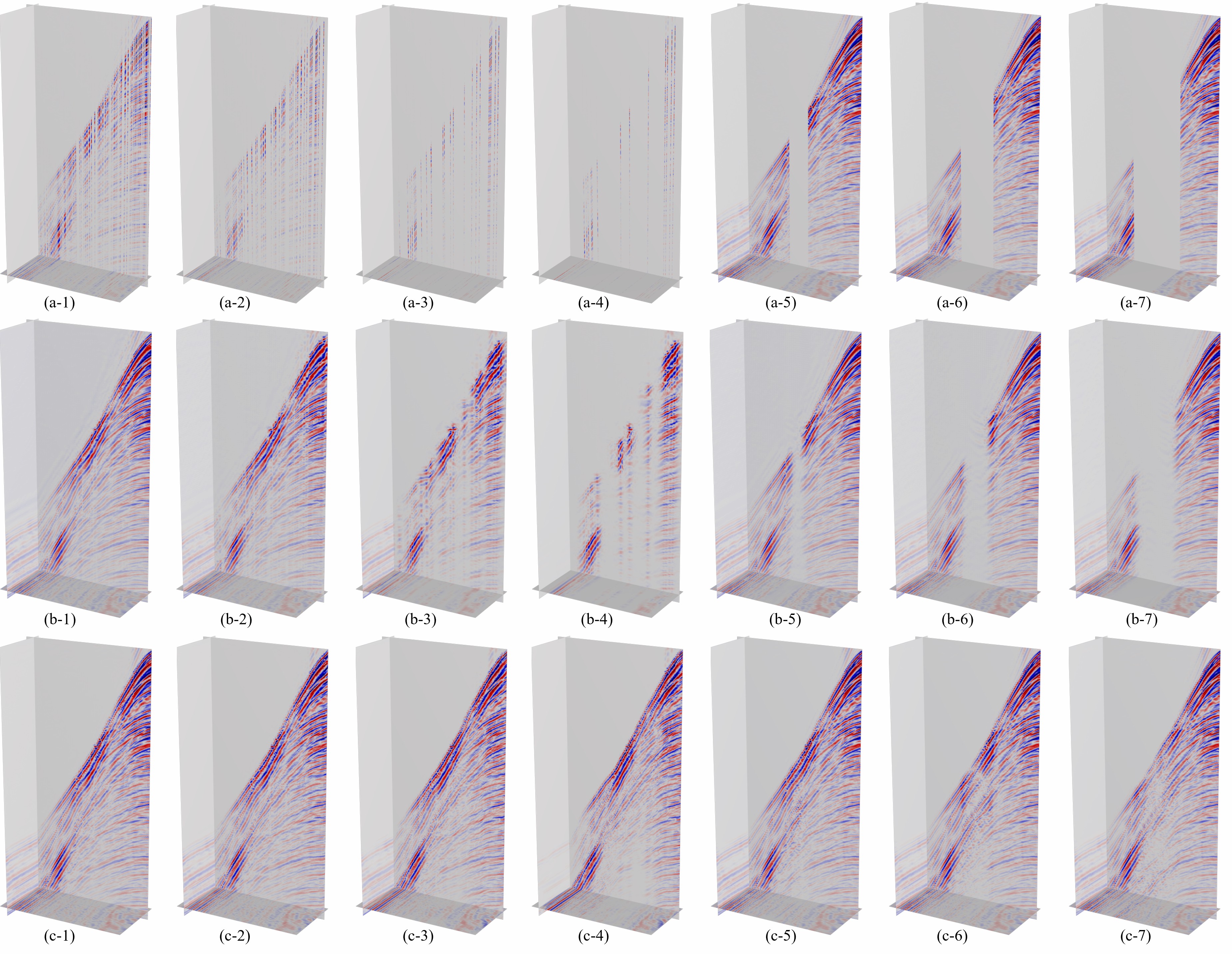}
	\centering\caption{
		(a-1)-(a-4) correspond to 50\%, 75\%, 90\% and 95\% discrete missing, respectively, (a-5)-(a-7) correspond to 40, 70 and 100 consecutive missing, respectively.
		(b-1)-(b-7) are the interpolation results of UNet for various cases.
		(c-1)-(c-7) are the interpolation results of MDA GAN for various cases.
		}
	\label{exp2}
\end{figure*}

Fig. \ref{exp2} shows the qualitative interpolation results of MDA GAN and UNet, and Table \ref{tab:exp3} shows the corresponding quantitative results.

At 50\% discrete deficiency, both MDA GAN and UNet are able to perform the interpolation task well. When the ratio comes to 75\%, UNet starts to be overwhelmed, while our method still shows high reliability.
With more complex 90\% and 95\% missing, UNet invalidates, while our method still yields promising results.
In the case of consecutive missing cases, all interpolation results of UNet are failures.
Fig. \ref{exp2} shows that the qualitative results of our method in complex cases are significantly higher than those of UNet, which is currently the most used, and the same conclusion can be obtained in the quantitative results in Table \ref{tab:exp3}.

\begin{table}[]
\centering
\caption{SSIM and PSNR on Mobil Avo Viking Graben Line 12}
\label{tab:exp3}
	\begin{tabular}{@{}lccccl@{}}
		\toprule
		&
		\textit{\textbf{\begin{tabular}[c]{@{}c@{}}Discrete\\  Missing Ratio\end{tabular}}} &
		\textit{\textbf{50\%}} &
		\textit{\textbf{75\%}} &
		\textit{\textbf{90\%}} &
		\multicolumn{1}{c}{\textit{\textbf{95\%}}} \\ \midrule
		\textit{\textbf{SSIM}} & UNet          & 0.9677 & 0.9531 & 0.9060 & \multicolumn{1}{c}{0.8583} \\
		\textit{\textbf{}}     & MDA GAN (TCE) & 0.9885 & 0.9766 & 0.9419 & \multicolumn{1}{c}{0.9088} \\ \midrule
		\textit{\textbf{PSNR}} & UNet          & 35.09  & 30.14  & 23.76  & \multicolumn{1}{c}{22.00}  \\
		\textit{\textbf{}}     & MDA GAN (TCE) & 36.11  & 31.59  & 27.44  & \multicolumn{1}{c}{25.93}  \\ \midrule
		&
		\textit{\textbf{\begin{tabular}[c]{@{}c@{}}Continuous \\ Missing Traces\end{tabular}}} &
		\textit{\textbf{40}} &
		\textit{\textbf{70}} &
		\textit{\textbf{100}} &
		\\ \midrule
		\textit{\textbf{SSIM}} & UNet          & 0.9627 & 0.9541 & 0.9349 &                            \\
		& MDA GAN (TCE) & 0.9845 & 0.9682 & 0.9510 &                            \\ \midrule
		\textit{\textbf{PSNR}} & UNet          & 35.02  & 28.82  & 28.43  &                            \\
		& MDA GAN (TCE) & 35.19  & 30.33  & 29.24  &                            \\ \bottomrule
	\end{tabular}
\end{table}

\subsection{Test on New Zealand Kerry and F3 Netherlands}
This work focuses on the interpolation and reconstruction of complex cases. The above experiments validate the reconstruction performance of our method under complex missing for the set of prestack gather. The purpose of this subsection is to verify the performance of our method with complex surveys rich in anomalies such as faults and salt bodies.

The missing in original engineering often occurs in shot gathers, and anomalous bodies also appear in brutal stack sections \cite{niu2021seismic} formed by shot gathers, but the post-stack data containing anomalous bodies is more abundant in the current public data, so we use the post-stack as the test data in this subsection.
\subsubsection{New Zealand Kerry (fault-rich)}
\begin{figure*}
	\centering
	\includegraphics[scale=0.58]{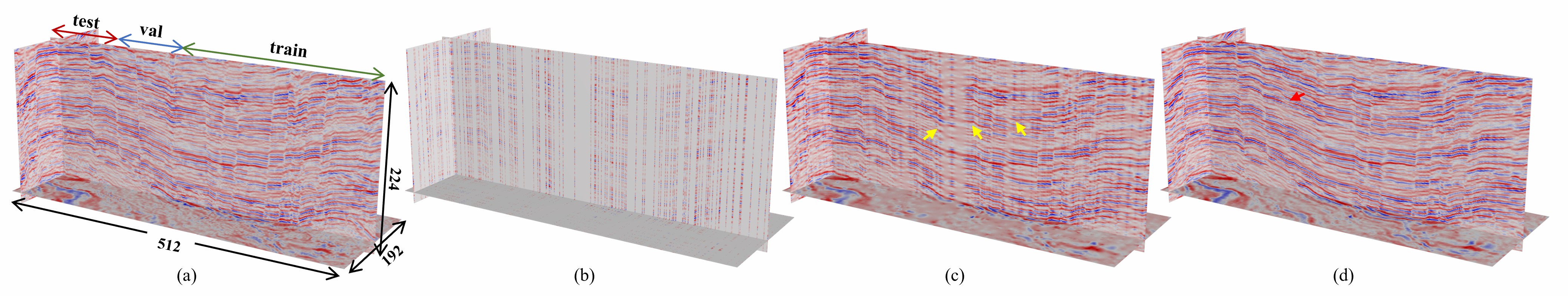}
	\centering\caption{
	(a) New Zealand Kerry original data, (b) 80\% traces loss in both inline and crossline directions, (c) UNet interpolation results, (d) MDA GAN interpolation results.
	}
	\label{exp4}
\end{figure*}

Kerry contains abundant faults perpendicular to the timeline, which is a great challenge for the interpolation task.
Fig. \ref{exp4} (a) shows the original data, which sampled at a total of $224 \times 512 \times 192$ grid points. Due to the limited data available for Kerry, part of the data used for the presentation was involved in the training (which has been marked in the Fig. \ref{exp4} (a)).

In Fig. \ref{exp4} (b), we removed 80\% of slices in both crossline and timeline directions, respectively, accumulating 95.62\% of the missing voxels. Because of the large proportion of missing traces, some areas form continuous missing, and most of the faults are covered by missing traces.

UNet does not complete the reconstruction for most of the missing faults, and there are some voids at the yellow arrows in (c), while at the corresponding positions, MDA GAN fills these voids as faults. In the region marked by the red arrow in (d), although MDA GAN restores the amplitude, the fault information is lost because the continuous missing here removes the fault completely, while MDA GAN shows a very high interpolation performance for the missing where a portion of the fault information is retained.

\subsubsection{F3 Netherlands (salt bodies)}
Salt bodies are common in marine seismic data, which have closed characteristics and distinct boundaries. We used the salt body part at the bottom of the F3 as qualitative data, which was all used as validation and test set without involvement in training, which sampled at a total of $128 \times 384 \times 384$ grid points.

Fig. \ref{exp4} (a) shows the original data. In (b), we removed 80\% of slices in both crossline and timeline directions, respectively, accumulating 95.76\% of the missing voxels. In (c), the internal structure of the salt body is lost at (1) of the UNet reconstruction result, and the boundary information of the salt body is not reconstructed at (2), while the corresponding position in (d) shows a high-quality reconstruction result.
MDA GAN demonstrates excellent reconstruction performance for anomalies such as faults and salt bodies.

\begin{figure*}
	\centering
	\includegraphics[scale=0.6]{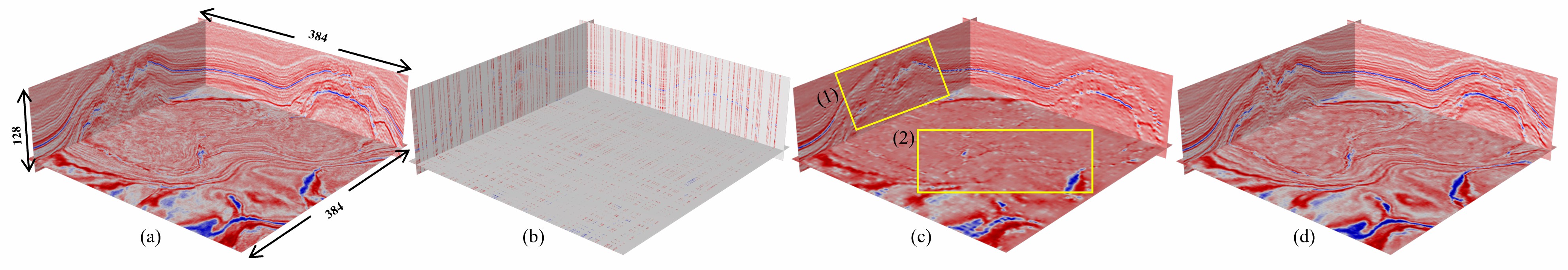}
	\centering\caption{
(a) F3 Netherlands original data, (b) 80\% traces loss in both inline and crossline directions, (c) UNet interpolation results, (d) MDA GAN interpolation results.
	}
	\label{exp5}
\end{figure*}

\subsection{Quantitative metrics for the total test set}

\begin{figure}
	\centering
	\includegraphics[scale=0.32]{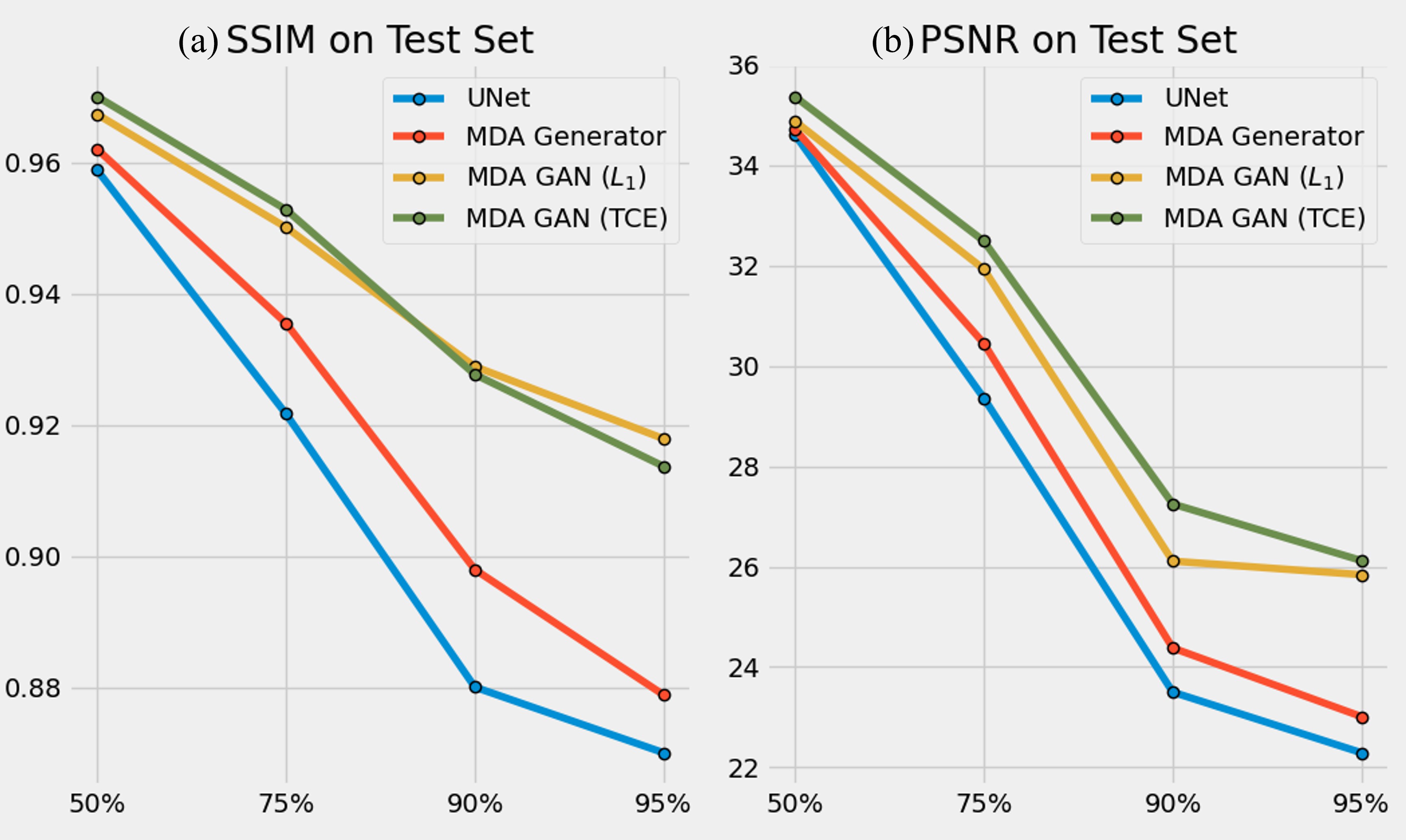}
	\centering\caption{
	(a) SSIM on test set (b) PSNR on test set
	}
	\label{exp6}
\end{figure}

Finally we calculate the SSIM and PSNR metrics jointly for all the test sets shown in Table \ref{datausage}. Because the size of the test parts differs for each survey, we calculate only the proportional discrete missing cases, the results are shown in Table \ref{tab:exp4}. For a more visual analysis of the performance of each group of experiments, we organized Table \ref{tab:exp4} into Fig. \ref{exp6}.

In Fig. \ref{exp6} (a), the MDA-based metrics are remarkably higher than the other two groups, indicating the effectiveness of multidimensional adversarial. While using $L_1$ and TCE losses do not differ significantly in terms of SSIM, with $L_1$ slightly higher than TCE, which is one of the reasons why L1 is currently widely used as a reconstruction loss, but the experiments from above show that the qualitative results of TCE losses are more advanced (Fig. \ref{exp1} and \ref{exp3}), and it can provide smoother and continuous reconstruction results.
In (b), the PSNR of MDA GAN (TCE) was higher than the other groups. Similar to (a), there was no significant difference between TCE and $L_1$, and TCE was slightly higher than $L_1$ in terms of PSNR metrics. This may be because PSNR is more sensitive to pixel differences, while SSIM prefers to show structural differences in regions.

\begin{table}[]
	\centering
	\caption{SSIM and PSNR on Mobil Avo Viking Graben Line 12}
	\label{tab:exp4}
	\begin{tabular}{@{}lccccc@{}}
		\toprule
		&
		\textit{\textbf{\begin{tabular}[c]{@{}c@{}}Discrete\\ Missing Ratio\end{tabular}}} &
		\textit{\textbf{50\%}} &
		\textit{\textbf{75\%}} &
		\textit{\textbf{90\%}} &
		\textit{\textbf{95\%}} \\ \midrule
		\textit{\textbf{}}     & UNet            & 0.9588 & 0.9217 & 0.8801 & 0.8701 \\
		\textit{\textbf{SSIM}} & MDA Generator   & 0.9620 & 0.9355 & 0.8980 & 0.8789 \\
		\textit{\textbf{}}     & MDA GAN ($L_1$) & 0.9673 & 0.9501 & 0.9289 & 0.9179 \\
		\textit{\textbf{}}     & MDA GAN (TCE)   & 0.9700 & 0.9528 & 0.9277 & 0.9137 \\ \midrule
		\textit{\textbf{}}     & UNet            & 34.61  & 29.36  & 23.50  & 22.29  \\
		\textit{\textbf{PSNR}} & MDA Generator   & 34.72  & 30.45  & 24.38  & 23.01  \\
		\textit{\textbf{}}     & MDA GAN ($L_1$) & 34.88  & 31.93  & 26.11  & 25.84  \\
		\textit{\textbf{}}     & MDA GAN (TCE)   & 35.37  & 32.49  & 27.25  & 26.12 \\ \bottomrule
	\end{tabular}
\end{table}

\section{Conclusion}

This work focuses on complex case interpolation of 3D seismic data, and proposes a multidimensional adversarial (MDA) seismic reconstruction framework and TCE reconstruction loss. The MDA framework emphasizes maintaining the anisotropy and spatial continuity of the reconstructed seismic data in all three directions, and thus promising results can be obtained in multiple complex cases.
The feature splicing module (FSM) embedded in the generator enables the reconstructed data to retain more information about the unmissing parts.
The TCE loss is mathematically derived to provide the generator with the optimal reconstruction gradient under the Tanh activation function to ensure that the reconstructed pixels are free of distortion.
Qualitative experiments have demonstrated the effectiveness of the MDA framework by showing that the MDA GAN is able to accomplish promising reconstructions under large proportions of discrete and continuous missing as well as fault and salt enriched data missing.
With SSIM visualization, it is demonstrated that the FSM module is indeed able to retain more of the original input information. Multiple qualitative experiments also demonstrate the ability to generate smoother and more continuous reconstruction results with Tanh cross entropy (TCE) loss.
Quantitative experiments demonstrated significantly better reconstruction performance of MDA GAN over multiple surveys than the baseline model and the ablation comparison experimental group. In the test set consisting of multiple surveys, MDA GAN outperformed the baseline model by 0.03, 0.05, and 0.04 for SSIM metrics and 3.13, 3.75, and 3.83 for PSNR under large proportional deletions of 75\%, 90\%, and 95\%, respectively. In the 50\% simpler missing case, our method still has the advantage.

\section{Appendix}\label{appendix}
\subsection{Generative Adversarial Network}
Generative Adversarial Network (GAN) are proposed by Goodfellow et al.\cite{goodfellow2014generative}. It consists of a generator network $\mathcal{G}_{\theta_G}$ and a discriminator network $\mathcal{D}_{\theta_D}$, where the task of the generator is to yield images $x\in\mathcal{R}^{N_w\times N_h}$ with a latent noise prior vector, $z\in\mathcal{R}^d$ as input $z$ is sampled from a known distribution, i.e. latent vector $z$, $z\sim \mathcal{U}[-1,1]^d$ \cite{goodfellow2014generative}. The task of the discriminator is to distinguish real images from generated ones. The generator and the discriminator play a zero-sum game in which the two networks learn from each other and the data obtained by the generator becomes increasingly close to the real data, so that the desired data can be generated. The objective function of GANs can be expressed by the equation (\ref{gans}).
\begin{equation}
	\begin{aligned}
		&\mathop{\text{max}}\limits_{\mathcal{G}_{\theta_G}}\mathop{\text{min}}\limits_{\mathcal{D}_{\theta_D}}V(\mathcal{G}_{\theta_G},\mathcal{D}_{\theta_D})=
		\\ & \mathbb{E}_{x\sim p_{data}(x)}[\text{log}\mathcal{D}_{\theta_D}(x)] +\mathbb{E}_{z\sim p_{z}(z)}[ \text{log}(1-\mathcal{D}_{\theta_D}(\mathcal{G}_{\theta_G}(z))]
	\end{aligned} \label{gans}
\end{equation}
where $p_{data}(x)$ represents the distribution of the training set and $p_{z}(z)$ represents the normal distribution of the noise, $\mathcal{D}_{\theta_D}(\mathcal{G}_{\theta_G}(z))$  represents the generated data.
$\mathcal{G}_{\theta_G}$ expects $\mathcal{D}_{\theta_D}(\mathcal{G}_{\theta_G}(z))$  to be as large as possible, when $V(\mathcal{G}_{\theta_G},\mathcal{D}_{\theta_D})$ becomes smaller, so equation is minimized for $\mathcal{G}_{\theta_G}$. $\mathcal{D}_{\theta_D}(x)$ is enhanced with the discriminator performance, when $\mathcal{D}_{\theta_D}(\mathcal{G}_{\theta_G}(z))$ becomes smaller and $V(\mathcal{G}_{\theta_G},\mathcal{D}_{\theta_D})$ becomes larger, so equation is maximized for $\mathcal{D}_{\theta_D}$.

\subsection{Image Inpainting via GAN} \label{inpvg}
\begin{figure}[htb]
	\centering
	\includegraphics[scale=0.52]{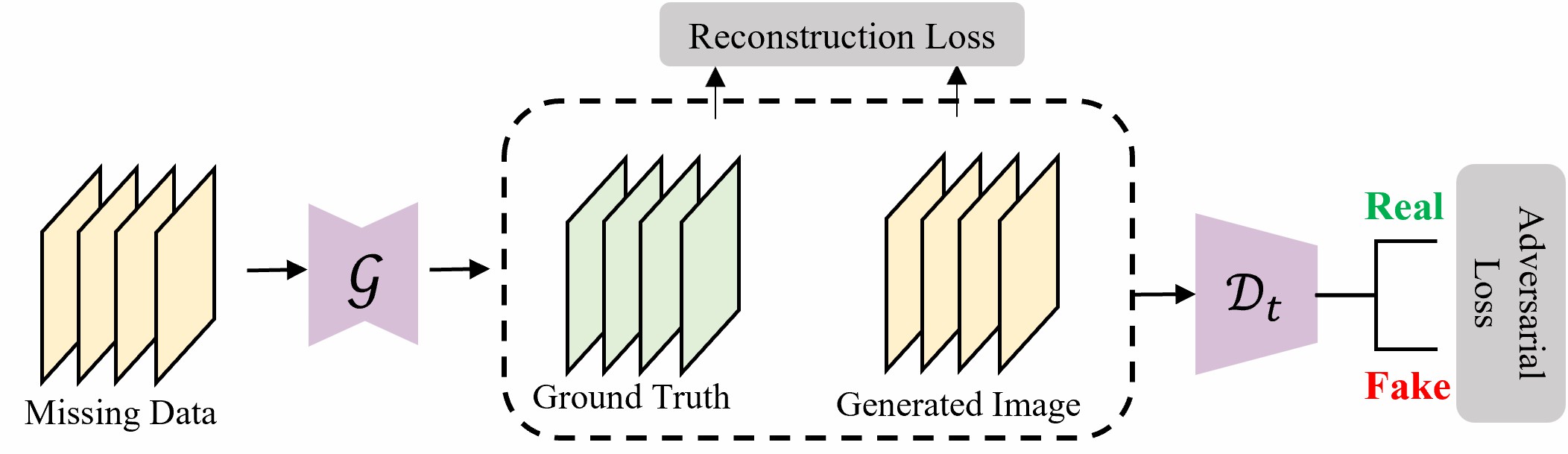}
	\centering\caption{The baseline framework for image inpainting based on GAN. Its network consists of a self-encoder and a discriminator, the loss consists of reconstruction loss and adversarial loss.}
	\label{gan4inp}
\end{figure}

The key to applying GAN to image inpainting is to join discriminator and adversarial loss on the base of encoder regression \cite{newell2016stacked,badrinarayanan2017segnet,ronneberger2015u,noh2015learning}. The existing image inpainting work is for 2D images, its loss can be expressed as equation (\ref{gan4inploss}),  the basic framework is shown in Fig \ref{gan4inp}.
\begin{equation}
	\mathcal{L}_\mathcal{G} = \mathcal{L}_{\text{adv}}+ \lambda\mathcal{L}_\text{rec} \label{gan4inploss}
\end{equation}
where $\mathcal{L}_{\text{adv}}$ is adversarial loss, commonly used CE-GAN\cite{goodfellow2014generative}, LS-GAN \cite{qi2020loss}, W-GAN\cite{arjovsky2017wasserstein}, etc., $\lambda\mathcal{L}_\text{rec}$ is the reconstruction loss (regression loss), $L_1$ and $L_2$ are generally used \cite{newell2016stacked,badrinarayanan2017segnet,ronneberger2015u,noh2015learning}.

Most of the current GAN-based image inpainting follows this framework.

\section*{Acknowledgment}
The authors are very indebted to the anonymous referees for their critical comments and suggestions for the improvement of this paper.

\bibliographystyle{IEEEtran}
\bibliography{references}

\end{document}